\documentclass[12pt]{article}

\def\bfkappa{\mathop{\mbox{\boldmath $\kappa$}}}

\newcommand{\be}{\begin{equation}}
\newcommand{\ee}{\end{equation}}
\newcommand{\bi}[1]{\vspace{-3mm} \bibitem{#1}}

\usepackage{epsfig,amsmath}
\usepackage{graphics}

\begin{document}

\begin{center}
Journal of Physics A. Vol.39. No.26. (2006) pp.8395-8407.
\end{center}

\begin{center}
{\Large \bf Psi-Series Solution of Fractional Ginzburg-Landau Equation}
\vskip 5 mm

{\large \bf Vasily E. Tarasov} \\

\vskip 3mm

{\it Skobeltsyn Institute of Nuclear Physics, \\
Moscow State University, Moscow 119992, Russia } \\
{E-mail: tarasov@theory.sinp.msu.ru}
\end{center}

\vskip 11 mm

\begin{abstract}
One-dimensional Ginzburg-Landau equations
with derivatives of noninteger order are considered.
Using psi-series with fractional powers, the solution of 
the fractional Ginzburg-Landau (FGL) equation is derived.
The leading-order behaviours of solutions about an arbitrary 
singularity, as well as their resonance structures, 
have been obtained.
It was proved that fractional equations of order $\alpha$ 
with polynomial nonlinearity of order $s$  
have the noninteger power-like behavior of order $\alpha/(1-s)$ 
near the singularity.
\end{abstract}



\section{Introduction}

Differential equations that contain derivatives 
of noninteger order \cite{SKM,OS} are called fractional equations 
\cite{Podlubny,KST}. 
The interest to fractional equations has been growing continually 
during the last few years because of numerous applications. 
In a fairly short period of time the areas of 
applications of fractional calculus have become broad.
For example, the derivatives and integrals of fractional order 
are used in chaotic dynamics \cite{Zaslavsky1,Zaslavsky2},
material sciences \cite{Hilfer,C2,Nig1}, 
mechanics of fractal and complex media \cite{Mainardi,Media},
quantum mechanics \cite{Laskin,Naber}, 
physical kinetics \cite{Zaslavsky1,Zaslavsky7,SZ,ZE},
plasma physics \cite{CLZ,Plasma2005}, 
astrophysics \cite{CMDA},
long-range dissipation \cite{GM}, 
non-Hamiltonian mechanics \cite{nonHam,FracHam},
long-range interaction \cite{LZ,TZ3,KZT}, 
anomalous diffusion and transport theory 
\cite{Zaslavsky1,Montr,Uch}.

The fractional generalization of the Ginzburg-Landau equation
was suggested in \cite{Zaslavsky6}.
This equation can be used to describe the dynamical processes 
in continuums with fractal dispersion and 
the media with fractal mass dimension \cite{Physica2005,Mil,TZ2}. 
In this paper, we generalize the psi-series 
approach to the fractional differential equations.
As an example, we consider a solution of 
the fractional Ginzburg-Landau (FGL) equation. 
We derive the psi-series for the one-dimensional FGL equation.
The leading-order behaviours of solutions about an arbitrary 
singularity, as well as their resonance structures, 
have been obtained.

In section 2, we recall the appearance of the Ginzburg-Landau equation 
with fractional derivatives.
In section 3, the singular behavior of the FGL equation is considered. 
In section 4, we discuss the powers of series terms 
that have arbitrary coefficients that are called 
the resonances or Kovalevskaya exponents.
In section 5, we derive the psi-series and recurrence relations
for one-dimensional FGL equation with
rational order $\alpha$ ($1<\alpha <2$).
In section 6, the example of an FGL equation with order $\alpha=3/2$
is suggested. In section 7, the next to singular behaviour 
for arbitrary (rational or irrational) order is discussed.
Finally, a short conclusion is given in section 8.

\section{Fractional Ginzburg-Landau (FGL) equation}

Let us recall the appearance of the nonlinear parabolic equation 
\cite{Leon,Light,Kad,ZS}, and 
the FGL equation \cite{Zaslavsky6,Physica2005,TZ2}.
Consider wave propagation in some media and 
present the wave vector $\bf k$ in the form
\begin{equation}
{\bf k} = {\bf k}_0 + {\bfkappa} = {\bf k}_0 + {\bfkappa}_{\parallel}
+ {\bfkappa}_{\perp}, 
\label{eq:36}
\end{equation}
where ${\bf k}_0$ is the unperturbed wave vector and subscripts
$(\parallel ,\perp )$ are taken respectively to the direction of ${\bf k}_0$. 
A symmetric dispersion law $\omega = \omega (k)$ for
$\kappa \ll k_0$ can be written as
\be \label{PE1}
\omega (k)= \omega (|{\bf k}|) \approx 
\omega (k_0)+ \ v_g \ (|{\bf k}| - k_0 )+
{1 \over 2} v^{\prime }_g \ (|{\bf k}| - k_0 )^2 ,
\ee
where
\be \label{PE2}
v_g = \left(\frac{\partial\omega}{\partial k} \right)_{k=k_0} , \quad
v^{\prime}_g=\left(\frac{\partial^2\omega}{\partial k^2}\right)_{k=k_0} ,
\ee
and
\be \label{PE3}
|{\bf k}|=|{\bf k}_0 + {\bfkappa}|=
\sqrt{({\bf k}_0+\kappa_{\parallel})^2+\kappa^2_{\perp}} \approx
k_0+\kappa_{\parallel}+\frac{1}{2k_0}\kappa^2_{\perp}.
\ee
Substitution of (\ref{PE3}) into (\ref{PE1}) gives
\be
\omega (k) \approx \omega_0 + v_g\kappa_{\parallel} +
{v_g\over 2k_0} \kappa_{\perp}^2 +
\frac{v^{\prime}_g}{2} \kappa_{\parallel}^2 ,
\label{eq:37}
\ee
where $\omega_0=\omega(k_0)$. 
Expression (\ref{eq:37}) in the dual space ("momentum representation") 
corresponds to the following equation in the coordinate space:
\begin{equation}
i {\partial Z \over \partial t} =\omega_0 Z- 
i v_g {\partial Z \over \partial x}
 - {v_g \over 2k_0 } \Delta_{\perp} Z-
{v^{\prime}_g \over 2 } \Delta_{\parallel} Z
 \label{eq:38}
\end{equation}
with respect to the field $Z=Z(t,x,y,z)$, 
where $x$ is along ${\bf k}_0$, 
and we use the operator correspondence between the dual space 
and usual space-time:
\[
\omega (k)  \ \longleftrightarrow \
 i {\partial\over\partial t} , \quad
\kappa_{\parallel} \ \longleftrightarrow \ 
 -  i {\partial\over\partial x} , \] 
\be
{(\bfkappa}_{\perp})^2 \ \longleftrightarrow \ - \Delta_{\perp}=
- {\partial^2 \over\partial y^2 }-{\partial^2 \over\partial z^2 } , \quad
{(\bfkappa}_{\parallel})^2 \ \longleftrightarrow \ 
- \Delta_{\parallel}=- {\partial^2 \over\partial x^2 } .
\label{eq:39}
\ee
A generalization to the nonlinear case can be carried out similarly to
(\ref{eq:37}) through a nonlinear dispersion law dependence on the wave
amplitude:
\begin{equation} 
\omega = \omega (k,|Z|^2 ) \approx\omega (k,0) + b|Z|^2
 = \omega (|{\bf k}|) + b|Z|^2
 \label{eq:40}
\end{equation}
with some constant 
$b = \partial\omega (k, |Z|^2 )/\partial |Z|^2$ at $|Z| = 0$. 
In analogy to (\ref{eq:38}), we obtain from (\ref{eq:37}) and (\ref{eq:39}): 
\begin{equation}
i {\partial Z \over \partial t} =\omega(k_0) Z - 
i v_g {\partial Z \over \partial x}
 - {v_g \over 2k_0 } \Delta_{\perp} Z-
{v^{\prime}_g \over 2 } \Delta_{\parallel} Z+ b|Z|^2 Z .
 \label{eq:41}
\end{equation}
This equation is known as the nonlinear parabolic equation 
\cite{Leon,Light,Kad,ZS}. 
The change of variables from $(t,x,y,z)$ to $(t,x-v_gt,y,z)$ gives
\begin{equation}
-i {\partial Z \over \partial t}
 = {v_g \over 2k_0} \Delta_{\perp} Z +
{v^{\prime}_g \over 2} \Delta_{\parallel} Z 
- \omega(k_0) Z - b|Z|^2 Z 
 \label{eq:42}
\end{equation}
that is also known as the nonlinear Schr\"{o}dinger (NLS) equation.

Wave propagation in media with fractal properties can be easily
generalized by rewriting the dispersion law (\ref{eq:37}), (\ref{eq:40})
in the following way \cite{Zaslavsky6}:
\begin{equation}
 \omega (k,|Z|^2 )= \omega (k_0 ,0) + v_g \kappa_{\parallel} +
g_1 ({\bfkappa}_{\perp}^2 )^{\alpha /2} +
g_2 ({\bfkappa}_{\parallel}^2 )^{\beta /2}+ b|Z|^2,
\quad (1<\alpha, \beta <2)
 \label{eq:43}
\end{equation}
with new constants $g_1$, $g_2$.

Using the connection between Riesz fractional derivative 
and its Fourier transform \cite{SKM} 
\begin{equation}
(-\Delta_{\perp} )^{\alpha /2} 
\longleftrightarrow
({\bfkappa}_{\perp}^2 )^{\alpha /2} ,
\quad
(-\Delta_{\parallel} )^{\beta /2} 
\longleftrightarrow  
({\bfkappa}_{\parallel}^2 )^{\beta /2} ,
 \label{eq:44}
\end{equation}
we obtain from (\ref{eq:43})
\begin{equation}
i {\partial Z \over \partial t} =- iv_g {\partial Z \over \partial x}
+ g_1 (-\Delta_{\perp} )^{\alpha /2}  Z 
+ g_2 (-\Delta_{\parallel} )^{\beta /2}  Z 
+ \omega_0 Z + b|Z|^2 Z,
  \label{eq:45}
\end{equation}
where $Z=Z(t,x,y,z)$. 
By changing the variables from $(t,x,y,z)$ to $(t,\xi,y,z)$, 
$\xi=x-v_gt$, and using 
\be
(-\Delta_{\parallel} )^{\beta /2}=
\frac{\partial^{\beta} }{\partial |x|^{\beta}}=
\frac{\partial^{\beta} }{\partial |\xi|^{\beta}}, 
\ee
we obtain  from (\ref{eq:45}) equation
\begin{equation}
i {\partial Z \over \partial t} =
 g_1 (-\Delta_{\perp} )^{\alpha /2}  Z 
+ g_2 (-\Delta_{\parallel} )^{\beta /2}  Z 
+ \omega_0 Z + b|Z|^2 Z,
\label{eq:46}
\end{equation}
that can be called the fractional nonlinear parabolic equation.
For $g_2=0$ we get the
nonstationary FGL equation (fractional NLS equation)
suggested in \cite{Zaslavsky6}.
Let us comment on the physical structure of (\ref{eq:46}).
The first term on the right-hand side is related to 
wave propagation in media with fractal properties. 
The fractional derivative can also appear
as a result of long-range interaction \cite{LZ,TZ3,KZT}. 
Other terms on the right-hand-side of equations (\ref{eq:45}) 
and (\ref{eq:46}) correspond to  wave interaction due to
the nonlinear properties of the media. Thus, equation (\ref{eq:46}) 
can describe fractal processes of self-focusing and related issues.

We may consider one-dimensional simplifications of (\ref{eq:46}), i.e., 
\begin{equation}
i {\partial Z \over \partial t}=
g_2 \frac{\partial^{\beta} Z}{\partial |\xi|^{\beta}}
+ \omega_0 Z + b|Z|^2 Z,
\label{eq:47a}
\end{equation}
where $Z=Z(t,\xi)$, $\xi=x-v_gt$, or the equation
\begin{equation}
i {\partial Z \over \partial t} =
g_1 \frac{\partial^{\alpha} Z}{\partial |z|^{\alpha}}
+ \omega_0 Z + b|Z|^2 Z,
\label{eq:47b}
\end{equation}
where $Z=Z(t,z)$.
We can reduce (\ref{eq:47b}) to the case of a propagating wave 
\be
Z=Z(z-v_gt)\equiv Z(\eta) .
\ee
For the real field $Z$, Eq. (\ref{eq:47b}) becomes
\be \label{eq:49}
g_1\frac{d^{\alpha}Z}{d |\eta|^{\alpha}}+c\frac{dZ}{d\eta}+\omega_0 Z+bZ^3=0,
\quad \eta=z-v_g t ,
\ee
where $c=iv_g$.
This equation takes the form of the fractional generalization of
the Ginzburg-Landau equation, when $v_g=0$.

It is well known that the nonlinear term in equations of type
(\ref{eq:42})  leads to a steepening of the solution and its
singularity. The steepening process may be stopped by a diffusional or
dispersional term, i.e. by a higher derivative term. A similar phenomenon
may appear for the fractional nonlinear equations (\ref{eq:49}).

\section{Singular behavior of FGL equation}

There is an approach to the question of integrability which 
is not concerned with the display of explicit functions, 
but with the demonstration of a specific property. 
This is the existence of Laurent series for each of the dependent variables. 
The series may not be summable to an explicit form, but does represent 
an analytic function. The essential feature of this 
Laurent series is that it is an expansion about a particular type of
movable singularity, i.e., a pole. 
Consequently the existence of these Laurent series is intimately 
concerned with the singularity analysis of differential 
equations initiated about a century ago by Painleve, Gambier and Garnier 
\cite{Ince} and continued since by many workers
including Bureau \cite{3} and Cosgrove et al \cite{4}.

The connection of this type of singular behavior and 
the solution of partial differential equations by the method of the 
inverse scattering transform was noticed by Ablowitz et al \cite{Ablowitz}
who developed an algorithm, called the ARS algorithm, to test whether 
the solution of an ordinary differential equation was expressible 
in terms of a Laurent expansion. 
If this was the case, the ordinary differential equation 
was said to pass the Painleve test and was conjectured to be integrable. 
Under more precise conditions Conte \cite{8} has shown that 
the equation is integrable. Psi-series solutions of differential 
equations are considered in \cite{Tabor1,Tabor2,BSV,CTW}.

In the paper, we consider the fractional equation 
\be \label{FGL}
gD^{\alpha}_x Z(x)+cD^{1}_x Z(x)+aZ(x) +bZ^3(x)=0,
\ee
where $1<\alpha < 2$, and $D^{\alpha}_x$ is 
the fractional Riemann-Liouville derivative:
\be
D^{\alpha}_x Z(x)=\frac{1}{\Gamma (m-\alpha)} \frac{d^m}{d x^m}  
\int^x_{x_0}  dy \frac{Z(y)}{(x-y)^{\alpha-m+1}} . \ee
Here, $m$ is the first whole number greater than or equal to $\alpha$.
In our case $m=2$. 
We detect possible singular behavior in the solution of a 
differential equation by means of the leading-order analysis. 

To determine the leading-order behavior, we set
\be \label{Zfp}
Z(x)=f(x-x_0)^{p} ,
\ee
where $x_0$ is an arbitrary constant (the location of the singularity).  
Then, we substitute (\ref{Zfp}) into the fractional differential 
equation (\ref{FGL}) and look for two or more dominant terms.
The detection of which terms are dominant is identical to the determination
of which terms in an equation are self-similar.

Substituting (\ref{Zfp}) into equation (\ref{FGL}), and
using the relation
\be \label{Da}
D^{\alpha}_x x^p=
\frac{\Gamma(p+1)}{\Gamma(p+1-\alpha)} x^{p-\alpha} , \quad (p>-1) ,
\ee
we get
\be
gf \frac{\Gamma(p+1)}{\Gamma(p+1-\alpha)} (x-x_0)^{p-\alpha}+
cpf(x-x_0)^{p-1}+af (x-x_0)^p+ bf^3 (x-x_0)^{3p}=0.
\ee
If $1< \alpha < 2$, then $p-\alpha < p-1$. For the dominant terms, 
\be \label{dominant}
gf \frac{\Gamma(p+1)}{\Gamma(p+1-\alpha)} (x-x_0)^{p-\alpha}+
bf^3 (x-x_0)^{3p}=0.
\ee
As a result, we obtain
\be \label{p}
p-\alpha=3p,
\ee
\be \label{f}
g\frac{\Gamma(p+1)}{\Gamma(p+1-\alpha)} + bf^2 =0.
\ee
Equation (\ref{p}) gives
\be 
p=-\frac{\alpha}{2}.
\ee
If $1<\alpha<2$, then $-1<p<-1/2$. 
Therefore the leading-order singular behavior is found:
\be
Z(x)=f(x-x_0)^{-\alpha/2}, \quad 
f^2+\frac{g\Gamma(1-\alpha/2)}{b\Gamma(1-3\alpha/2)}=0,
\ee
and the singularity is a pole of order $\alpha/2$.
Evidently our psi-series starts at
$(x-x_0)^{-\alpha/2}$.
The resonance conditions and psi-series 
is considered in the next sections.

As a result, we get that
fractional differential equations of order $\alpha$ 
with polynomial nonlinearity of order $s$
have the noninteger power behavior of order $\alpha/(1-s)$ 
near the singularity.

\section{Resonance terms of FGL equation}

The powers of $(x-x_0)$ that have arbitrary coefficients
are called the resonances or Kovalevskaya's exponents. 
To find resonance, we consider the substitution
\be \label{Zfplr}
Z(x)=f (x-x_0)^p +l (x-x_0)^{p+r}, 
\ee
and find the values $r$. 
In equation (\ref{Zfplr}) parameters $p$ and $f$ are defined  by
\be \label{p-f0}
p=-\frac{\alpha}{2}, \quad
f=\sqrt{-\frac{g\Gamma(1-\alpha/2)}{b\Gamma(1-3\alpha/2)}} .
\ee
Substitution of Eq. (\ref{Zfplr}) into (\ref{FGL}) gives
\[ gf \frac{\Gamma(p+1)}{\Gamma(p+1-\alpha)} (x-x_0)^{p-\alpha}+
cpf(x-x_0)^{p-1}+af (x-x_0)^p+ bf^3 (x-x_0)^{3p}+\]
\[ +g l \frac{\Gamma(p+r+1)}{\Gamma(p+r+1-\alpha)} (x-x_0)^{p+r-\alpha}+
cpl(x-x_0)^{p+r-1}+al (x-x_0)^{p+r}+ bl^3 (x-x_0)^{3p+3r}+ \]
\be \label{29}
+3bl^2 f (x-x_0)^{2p+3r}+3bl f^2 (x-x_0)^{p+3r}=0 .
\ee
Using equation (\ref{p-f0}), and considering 
the linear with respect to $l$ terms of (\ref{29}), we have
\be \label{res}
\frac{\Gamma(1+r-\alpha/2)}{\Gamma(1+r-3\alpha/2)}-
3 \frac{\Gamma(1-\alpha/2)}{\Gamma(1-3\alpha/2)}=0 .
\ee
This relation allows us to derive the values of $r$.
Equation (\ref{res}) can be directly derived by using the recurrence relations.
In the general case, the values of $r$ can be irrational or complex numbers.
The solution of the FGL equation with $1<\alpha \le 2$ have 
two arbitrary parameters.
Therefore, we must have two values of $r$ that are the 
solutions of equation (\ref{res}). 
It is interesting to note that (\ref{res}) gives
two real values of $r$ only for 
\be
\alpha >\alpha_0 ,
\ee
where
\be
\alpha_0 \approx 1.3005888986 .
\ee
The order $\alpha_0$ is an universal value that does not depends on
values of parameters $g$, $a$, $b$, $c$ of the FGL equation (\ref{FGL}).

The plots of the function
\be
y(r)=\frac{\Gamma(1+r-\alpha/2)}{\Gamma(1+r-3\alpha/2)}-
3 \frac{\Gamma(1-\alpha/2)}{\Gamma(1-3\alpha/2)} 
\ee
are shown in figures 1 and 2.
The solutions of equation (\ref{res}) correspond to the points of 
intersections with the horizontal axis. 

As a result, the nature of the resonances is summarized as follows: \\
1) For $\alpha$ such that $1< \alpha < \alpha_0$ the values of $r$
are complex or $r <-\alpha/2$. \\
2) For $\alpha$ such that $\alpha_0 < \alpha <2$, we have two the real 
values of $r$. Note that for $\alpha_0 <\alpha <1.9999999995$,
the values $r$  satisfy the inequality $r<6.426$.


\begin{figure}[tbh]
\begin{center}
\resizebox{9cm}{!}{\includegraphics{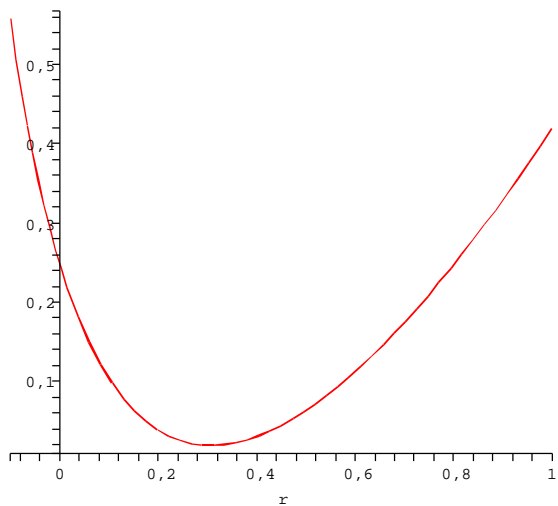}}
\caption{\it Plot for the order $\alpha=1.30$. }  
\label{res1}
\end{center}
\end{figure}
\begin{figure}[tbh]
\begin{center}
\resizebox{9cm}{!}{\includegraphics{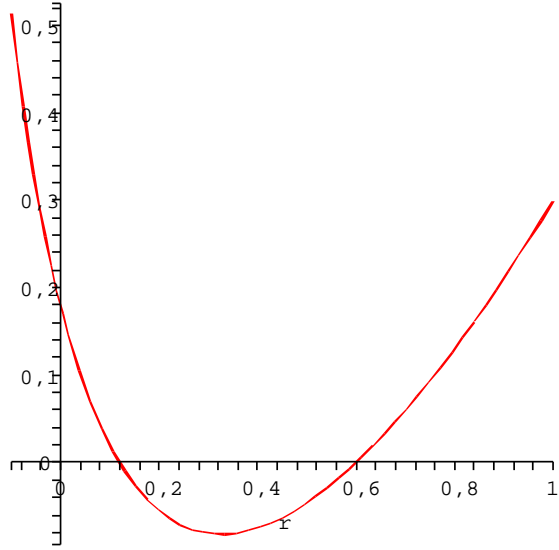}}
\caption{\it Plot for the order $\alpha=1.31$. }  
\label{res2}
\end{center}
\end{figure}


\section{Psi-Series for FGL Equation of Rational Order}

Let us consider the psi-series and recurrence relations
for the one-dimensional FGL equation (\ref{FGL}), 
where the order $\alpha$ is a rational number such that:
\be
\alpha=\frac{m}{n}, \quad (1<\alpha <2).
\ee

Following a standard procedure \cite{Tabor1}, we substitute
\be \label{Zsum1}
Z(x)=\frac{1}{(x-x_0)^{\alpha/2}}
\sum^{\infty}_{k=0} e_k \phi^k(x-x_0) ,
\ee
where 
\be
e_0=f=\sqrt{-\frac{g\Gamma(1-\alpha/2)}{b\Gamma(1-3\alpha/2)}}, 
\ee
into the fractional Ginzburg-Landau equation (\ref{FGL}).
Note that the coefficient $e_0$ is a real number for two cases: 
(1) $g/b\ge 0$ and $1<\alpha<4/3$; \ (2) $g/b\le0$ and $4/3 <\alpha <2$.

For the rational order $\alpha=m/n$, 
we suggest to use the $\phi$-function in the form
\be 
\phi(x-x_0)=(x-x_0)^{1/2n} .
\ee
Then 
\be \label{Zsum2}
Z(x)=\frac{1}{(x-x_0)^{\alpha/2}}
\sum^{\infty}_{k=0} e_k (x-x_0)^{\beta_k}=
\sum^{\infty}_{k=0} e_k (x-x_0)^{\beta_k-\alpha/2} ,
\ee
where 
\be
\beta_k=\frac{k}{2n} .
\ee
In this case, the action of fractional derivative of order $\alpha=m/n$ 
can be represented as 
the change of the number of term $e_k \rightarrow e_{k-2m}$:
\be
D^{\alpha}_x (x-x_0)^{\beta_k} =
\frac{\Gamma(\beta_k+1)}{\Gamma(\beta_{k-2m}+1)} (x-x_0)^{\beta_{k-2m}} .
\ee
It allows us to derive the generalized psi-series solutions
of fractional Ginzburg-Landau equation.

Substitution of the series
\be
Z(x)=\sum^{\infty}_{k=0} e_k (x-x_0)^{\beta_k-\alpha/2}=
\sum^{\infty}_{k=0} e_k (x-x_0)^{\frac{k-m}{2n}}
\ee
into equation (\ref{FGL}) gives
\[ g \sum^{\infty}_{k=0} e_k \frac{\Gamma(\frac{k-m+2n}{2n})}{\Gamma(\frac{k-3m+2n}{2n})}
(x-x_0)^{\frac{k-3m}{2n}}
+c \sum^{\infty}_{k=0} e_k \frac{k-m}{2n} (x-x_0)^{\frac{k-m-2n}{2n}}+ \]
\be \label{ser1}
 +a \sum^{\infty}_{k=0} e_k (x-x_0)^{\frac{k-m}{2n}}+
b \sum^{\infty}_{k_1=0} \sum^{\infty}_{k_2=0}\sum^{\infty}_{k_3=0} 
e_{k_1} e_{k_2} e_{k_3}  (x-x_0)^{\frac{k_1+k_2+k_3-3m}{2n}}=0. \ee
Let us compute $e_k$ ($k=1,2,...$) through the equation
of coefficients of like power of $(x-x_0)$ to zero in (\ref{ser1}):
\[ g \sum^{\infty}_{k=0} e_k 
\frac{\Gamma(\frac{k-m+2n}{2n})}{\Gamma(\frac{k-3m+2n}{2n})}
(x-x_0)^{\frac{k-3m}{2n}}
+c \sum^{\infty}_{k=2m-2n} e_{k-2m+2n} 
\frac{k-3m+2n}{2n} (x-x_0)^{\frac{k-3m}{2n}}+ \]
\[ +a \sum^{\infty}_{k=2m} e_{k-2m} (x-x_0)^{\frac{k-3m}{2n}}
+3b e^2_0 \sum^{\infty}_{k=0} e_k (x-x_0)^{\frac{k-3m}{2n}} +\]
\be \label{ser2} 
+b \sum^{\infty}_{k=0} \sum^{k-1}_{i=1} \sum^{i}_{j=1} 
e_{k-i-j} e_{i} e_{j}  (x-x_0)^{\frac{k-3m}{2n}}=0. \ee
Using $e^2_0=f^2$, we get
\[ g e_k \frac{\Gamma(\frac{k-m+2n}{2n})}{\Gamma(\frac{k-3m+2n}{2n})}
+c e_{k-2m+2n} \frac{k-3m+2n}{2n} + \]
\be \label{ser3}
+a e_{k-2m} +3b f^2 e_k +
b \sum^{k-1}_{i=1} \sum^{i}_{j=1} e_{k-i-j} e_{i} e_{j} =0. \ee
Here $k=0,1,2,...$, and $e_k=0$ for $k<0$.
We can rewrite the recurrence relation (\ref{ser3}) as
\be \label{r1} 
e_k \left(g\frac{\Gamma(\frac{k-m+2n}{2n})}{\Gamma(\frac{k-3m+2n}{2n})}
+ 3b f^2 \right)=
-c e_{k-2m+2n} \frac{k-3m+2n}{2n} -a e_{k-2m} 
-b \sum^{k-1}_{i=1} \sum^{i}_{j=1} e_{k-i-j} e_{i} e_{j} , \ee
where 
\be \label{f2}
f^2=-\frac{g\Gamma(1-\alpha/2)}{b\Gamma(1-3\alpha/2)}=
-\frac{g\Gamma(\frac{2n-m}{2n})}{b\Gamma(\frac{2n-3m}{2n})} .
\ee
Substitution of equation (\ref{f2}) into equation (\ref{r1}) gives 
\[ e_k g \left(\frac{\Gamma(\frac{k-m+2n}{2n})}{\Gamma(\frac{k-3m+2n}{2n})}
-\frac{3\Gamma(\frac{2n-m}{2n})}{\Gamma(\frac{2n-3m}{2n})} \right)=
-c e_{k-2m+2n} \frac{k-3m+2n}{2n} -a e_{k-2m}- \]
\be
-b \sum^{k-1}_{i=1} \sum^{i}_{j=1} e_{k-i-j} e_{i} e_{j} . \ee
Note that we get resonances for the $k$ that satisfies the condition
\be
\frac{\Gamma(\frac{k-m+2n}{2n})}{\Gamma(\frac{k-3m+2n}{2n})} 
-\frac{3\Gamma(\frac{2n-m}{2n})}{\Gamma(\frac{2n-3m}{2n})}=0. 
\ee
In this case, the coefficient $e_k$ can be arbitrary.

As a result,  we obtain the nonresonance terms
\be \label{mrec}
e_k= -A(k,m,n)
\left(c e_{k-2m+2n} \frac{k-3m+2n}{2n} +a e_{k-2m} +b 
\sum^{k-1}_{i=1} \sum^{i}_{j=1} e_{k-i-j} e_{i} e_{j} \right) ,
\ee
where
\be \label{mrec2}
A(k,m,n)= - \frac{ \Gamma(\frac{k-3m+2n}{2n}) \Gamma(\frac{2n-3m}{2n}) }{g 
[ \Gamma(\frac{k-m+2n}{2n})\Gamma(\frac{2n-3m}{2n}) 
-3 \Gamma(\frac{k-3m+2n}{2n}) \Gamma(\frac{2n-m}{2n}) ]} .
\ee

\section{Fractional Ginzburg-Landau Equation with $\alpha=1.5$}

Let us consider the FGL equation (\ref{FGL}) with derivative
of order $\alpha=3/2$.
In this case, $n=2$, $m=3$, 
and the coefficients $e_k$ are defined by 
\be \label{e_k}
e_k= -\frac{\left(c e_{k-2} \frac{k-5}{4} +a e_{k-6} +b 
\sum^{k-1}_{i=1} \sum^{i}_{j=1} e_{k-i-j} e_{i} e_{j}\right) 
\Gamma(\frac{k-5}{4}) \Gamma(\frac{-5}{4}) }{
g [ \Gamma(\frac{k+1}{4})\Gamma(\frac{-5}{4}) 
-3 \Gamma(\frac{k-5}{4}) \Gamma(\frac{1}{4})]} . \ee

For $k=0$, we have
\be \label{e0}
e_0=f=\sqrt{ -\frac{g\Gamma(\frac{1}{4})}{b\Gamma(\frac{-5}{4})} } ,
\ee
where $\Gamma(\frac{1}{4})/\Gamma(\frac{-5}{4})>0$, and 
we suppose $g/b<0$.
For $k=1$, equation (\ref{e_k}) leads to $e_1=0$.
For $k=2$,
\be \label{k=2}
e_2= \frac{3 c e_{0}   
\Gamma(\frac{-3}{4}) \Gamma(\frac{-5}{4}) }{
4 g [ \Gamma(\frac{3}{4})\Gamma(\frac{-5}{4})
-3 \Gamma(\frac{-3}{4}) \Gamma(\frac{1}{4})]} . \ee
Using the relation (\ref{e0}), we get 
\be \label{e2}
e_2=-\frac{c \sqrt{-5(g/b)} \pi^{3/2} 2^{3/4} \Gamma(3/4) }{2g 
[2 \Gamma^4(3/4)+5\pi^2]} .
\ee
For $k=3$, $e_3=0$. For $k=4$, 
\be \label{k=4}
e_4= -\frac{\left(c e_{2} \frac{-1}{4}+b e_{0} e^2_{2} \right)
\Gamma(\frac{-1}{4}) \Gamma(\frac{-5}{4}) }{g 
[ \Gamma(\frac{5}{4})\Gamma(\frac{-5}{4}) 
-3 \Gamma(\frac{-1}{4}) \Gamma(\frac{1}{4}) ]} . \ee
Substitution of (\ref{e0}), and (\ref{e2}) into (\ref{k=4}) gives 
\be \label{e4}
e_4=\frac{c^2 \sqrt{5g\pi \sqrt{2}/b} 
\Gamma^7(3/4)}{4g^2[2\Gamma^4(3/4)+5\pi^2]^2} 
\ee
For $k=5$, we have $e_5=0$. For $k=6$, equation (\ref{e_k}) is
\be \label{k=6}
e_6= -\frac{\left(c e_{4} \frac{1}{4} +
a e_{0} +b e_2^3 +2b e_0 e_2 e_4 \right) 
\Gamma(\frac{1}{4}) \Gamma(\frac{-5}{4}) }{g 
[ \Gamma(\frac{7}{4})\Gamma(\frac{-5}{4}) 
-3 \Gamma(\frac{1}{4}) \Gamma(\frac{1}{4})]} . \ee
Substituting $e_0$, $e_2$ and $e_4$ from 
equations (\ref{e0}), (\ref{e2}), and (\ref{e4}) into (\ref{k=6}), 
we have
\[ e_6=-\frac{\sqrt{-5g/b} \pi^{3/2} 2^{3/4} \Gamma(3/4) }{3g^2 
[2\Gamma^4(3/4)+5\pi^2]^3 [2 \Gamma^4(3/4)-5\pi^2]} \Bigl(
c^3\Gamma^{12}(3/4)+5c^3\Gamma^8(3/4) \pi^2+ \]
\be \label{e6}
+20c^3 \Gamma^4(3/4) \pi^4+16a g^2 \Gamma^{12}(3/4) 
+120ag^2\Gamma^{8}(3/4) \pi^2 +300ag^2\Gamma^{4}(3/4) 
\pi^4+250ag^2\pi^6 \Bigr) . 
\ee

As the result, we obtain 
\[ Z(x)=\sum^{\infty}_{k=0} e_k (x-x_0)^{\frac{k-3}{4}}= \]
\be \label{coe}
=e_0(x-x_0)^{-3/4}+e_2(x-x_0)^{-1/4}+e_4(x-x_0)^{1/4}+e_6(x-x_0)^{3/4}+...
\ee
This is a power psi-series that present the solution of 
FGL equation of order $\alpha=1.5$.
The coefficients in equation (\ref{coe}) are defined by equations 
(\ref{e0}), (\ref{e2}), (\ref{e4}), and (\ref{e6}).
For example, $a=-b=c=g=1$ gives
\[ e_0 \approx 0.9615539375 , \quad e_2 \approx -0.2382246293, \] 
\[ e_4 \approx 0.001685563496 ,\quad e_6 \approx 0.3872134448 .\]

\section{Next to Singular Behavior}

In sections 5 and 6, we consider the rational $\alpha$. 
In this section, we suppose that the order $\alpha$ 
is an arbitrary (rational or irrational). 
Instead of imposing a series commencing at the power 
indicated by the singularity found by the leading-order analysis, 
we can determine the next to singular behavior by writing
\be \label{ZfpG}
Z(x)=f (x-x_0)^p +G(x), 
\ee
where
\be \label{p-f}
p=-\frac{\alpha}{2}, \quad
f=\sqrt{-\frac{g\Gamma(1-\alpha/2)}{b\Gamma(1-3\alpha/2)}} .
\ee
We can always write $Z(x)$ in the form (\ref{ZfpG}). 
To make the process useful, we require that
the first term be the leading order term, i.e., 
\be \label{xG0}
(x-x_0)^{-p}G(x)=(x-x_0)^{\alpha/2}G(x) \rightarrow 0 \quad
if \quad (x-x_0) \rightarrow 0 . 
\ee 
Substituting (\ref{ZfpG}) into (\ref{FGL}), and using (\ref{Da}),
we have
\[ gf \frac{\Gamma(p+1)}{\Gamma(p+1-\alpha)} (x-x_0)^{p-\alpha}+
cpf(x-x_0)^{p-1}+af (x-x_0)^p+ bf^3 (x-x_0)^{3p}+\]
\be \label{66}
+gD^{\alpha}_x G(x)+cD^1_x G(x)+aG(x)+bG^3(x)
+3 bf^2 (x-x_0)^{2p} G(x)+ 3bf (x-x_0)^{p} G^2(x)=0 .
\ee
Equations (\ref{66}) and (\ref{p-f}) give
\[ gD^{\alpha}_x G(x)+cD^1_x G(x)+aG(x)+bG^3(x)
+3 bf^2 (x-x_0)^{2p} G(x)+ \]
\be
+3bf (x-x_0)^{p} G^2(x) cpf(x-x_0)^{p-1}+af (x-x_0)^p=0 .
\ee
Multiplying this equation on $(x-x_0)^{-3p}$,
and using condition (\ref{xG0}), we get
the equation without nonlinear terms for $(x-x_0)\rightarrow 0$.

As the result, we obtain
\be \label{linG}
gD^{\alpha}_x G(x)+cD^1_x G(x)+aG(x)+
cpf(x-x_0)^{-\alpha/2-1}+af (x-x_0)^{-\alpha/2}=0 ,
\ee
with condition (\ref{xG0}) for the solutions.
Equation (\ref{linG}) is a linear inhomogeneous fractional equation.
The solution of this equation allows us to find the
solution of the one-dimensional FGL equation.

Let us consider equation (\ref{linG}) with $c=0$, and the 
boundary conditions
\be
\left(D^{\alpha-k}_x G(x)\right)_{x=x_0}=G_k, \quad k=1, 2.
\ee
Then the solution is
\[ G(x)=\sum^{2}_{k=1} G_k (x-x_0)^{\alpha-k} 
E_{\alpha,\alpha+1-k} [-a (x-x_0)^{\alpha}]+ \]
\be \label{Gsol1}
+af \int^{x}_0 (x-x_0-y)^{\alpha-1} E_{\alpha, \alpha} [-a(x-x_0-y)^{\alpha}]
(y-x_0)^{-\alpha/2} dy .
\ee
Here $E_{\alpha,\beta}[z]$ is a Mitag-Leffler function that is defined by
\be
E_{\alpha,\beta}[z]=\sum^{\infty}_{k=0} \frac{z^k}{\Gamma(\alpha k+\beta)}.
\ee
Let us consider the integral representation 
for the Mittag-Leffler function
\be \label{58}
E_{\alpha,\beta}[z]=\frac{1}{2\pi i}\int_{Ha} 
\frac{\xi^{\alpha-\beta} e^{\xi}}{\xi^{\alpha}-z} d\xi ,
\ee
where $Ha$ denotes the Hankel path, 
a loop which starts from $-\infty$ along the lower side
of the negative real axis, encircles the circular disc 
$|\xi|\le |z|^{1/\alpha}$ in the positive direction, and 
ends at $-\infty$ along the upper side of the negative real axis.
By the replacement $\xi^{\alpha} \rightarrow \xi$ equation (\ref{58})
transforms into \cite{Podlubny,GLL}:
\be \label{59}
E_{\alpha,\beta}[z]=\frac{1}{2 \pi i \alpha} 
\int_{\gamma(a,\delta)} \frac{e^{ \xi^{1/\alpha} } 
\xi^{(1-\beta)/\alpha}}{\xi -z} d\xi, \quad (1<\alpha <2) ,
\ee
where $\pi \alpha/2 < \delta < min\{\pi, \pi \alpha\}$.
The contour $\gamma(a,\delta)$ consists of two rays
$S_{-\delta}=\{\arg(\xi)=-\delta, |\xi|\ge a\}$ and  
$S_{+\delta}=\{\arg(\xi)=+\delta, |\xi|\ge a\}$,
and a circular arc 
$C_{\delta}=\{|\xi|=1, -\delta \le arc(\xi) \le \delta \}$.
Let us denote the region on the left from $\gamma(a,\delta)$
as $G^{-}(a,\delta)$. Then \cite{GLL}:
\be \label{60}
E_{\alpha,\beta}[z]=-\sum^{\infty}_{n=1} 
\frac{z^{-n}}{\Gamma(\beta-\alpha n)}, \quad z \in G^{-}(a,\delta), 
\quad (|z| \rightarrow \infty),  
\ee 
and $\delta \le |\arg(z)|\le \pi$.
In our case, $z=-a(x-x_0)^{\alpha}$, $\arg(z)=\pi$.
As a result, we arrive at the asymptotic of the solution, 
which exhibits power-like tails for $x \rightarrow \infty$.
These power-like tails are the most important effect of 
the noninteger derivative in the fractional equations.

\section{Conclusion}

In this paper, we generalize the
psi-series approach to the fractional differential equations.
As an example, we consider the fractional Ginzburg-Landau (FGL) 
equation \cite{Zaslavsky6,Physica2005,Mil,TZ2}.
The suggested psi-series approach can be used for 
a wide class of fractional nonlinear equations. 
The leading-order behaviours of solutions about an arbitrary 
singularity, as well as their resonance structures, 
can be derived for fractional equations by the suggested 
generalization of psi-series.

In the paper, we use the series
\be \label{Con1}
Z(x)=\frac{1}{(x-x_0)^{m/2n}}
\sum^{\infty}_{k=0} e_k \; (x-x_0)^{k/2n} ,
\ee
where $k$, $m$, $n$ are the integer numbers. 
For the order $\alpha=m/n$, the action of fractional derivative 
\be
D^{\alpha}_x (x-x_0)^{k/2n} =
\frac{\Gamma(k/2n+1)}{\Gamma((k-2m)/2n+1)} (x-x_0)^{(k-2m)/2n} 
\ee
can be represented as 
the change of the number of term $e_k \rightarrow e_{k-2m}$ in (\ref{Con1}).
It allows us to derive the psi-series 
for the fractional differential equation of order $\alpha=m/n$.
For the FGL equation the leading-order singular behaviour
is defined by power that is equal to the half of derivative order. 

A remarkable property of the dynamics described by the equation with 
fractional space derivatives is that the 
solutions have power-like tails \cite{KZT}. 
In this paper, we prove that fractional differential 
equations of order $\alpha$ with a polynomial nonlinear 
term of order $s$ have the noninteger power-like behavior 
of order $\alpha/(1-s)$ near the singularity.

It is interesting to find barriers to integrability 
for fractional differential equations. 
In general, the integrability of fractional nonlinear equations 
is a very interesting object for future researches. 
It has many problems that are connected with specific 
properties of the fractional calculus.
For example, we must derive a generalization 
of the Lie algebra for the vector fields that are
defined by fractional derivatives.
For this generalization, the Jacobi identity cannot be satisfied, 
and we have a non-Lie algebra. 
The definition of such "fractional" Lie algebra is an open question 
and cannot be realized by a simple way. 
To formulate the fractional generalization of a Lie algebra 
for derivatives of noninteger oder, 
we can use the representation of fractional derivatives 
as infinite series of derivatives of integer orders \cite{SKM}.
For example, the Riemann-Liouville fractional derivative 
can be represented as 
\be
D^{\alpha}_x=\sum^{\infty}_{n=0} A_n(x,x_0,\alpha)
\frac{d^n}{dx^n} ,
\ee
where
\be
A_n(x,a,\alpha)= 
\frac{(-1)^{n-1}\alpha \Gamma(n-\alpha)}{\Gamma(1-\alpha) \Gamma(n+1)}
\frac{(x-x_0)^{n-\alpha}}{\Gamma(n+1-\alpha)}  .
\ee
Then the possible realization of the generalization is connected with 
the special algebraic structures for infinite jets \cite{Vin}.
These structures and approaches can help 
to solve some problems that are connected with
the integrability of fractional nonlinear equations.


\end{document}